\documentclass[traditabstract]{aa} 
\usepackage{graphicx}
\usepackage{txfonts}
\usepackage{natbib}
\bibpunct{(}{)}{;}{a}{}{,} 

\def\solar {\ifmmode_{\mathord\odot} \else $_{\mathord\odot}$ \fi}
\def\jup {\ifmmode_{\mathrm{Jup}} \else $_{\mathrm{Jup}}$ \fi}
\def\earth {\ifmmode_{\mathord\oplus} \else $_{\mathord\oplus}$ \fi}
\def\Msol {\ifmmode {\,\mathrm{M}\solar} \else \,M\solar \fi}     
\def\Rsol {\ifmmode {\,\mathrm{R}\solar} \else R\solar \fi}     
\def\Lsol {\ifmmode {\,\mathrm{L}\solar} \else L\solar \fi}     
\def\Mjup {\ifmmode {\,\mathrm{M}\jup} \else M\jup \fi}
\def\Mearth {\ifmmode {\,\mathrm{M}\earth} \else M\earth \fi}
\def\Rearth {\ifmmode {\,\mathrm{R}\earth} \else R\earth \fi}

\def\mps {\ifmmode {\,\mathrm{m\,s^{-1}}} \else $\mathrm{m\,s^{-1}}$ \fi}     

\begin{document}
  \title{A hot Uranus transiting the nearby M dwarf GJ3470\thanks{Based on observations made with the HARPS instrument on the ESO 3.6 m telescope under the program IDs 183.C-0437 at Cerro La Silla (Chile). Our radial-velocity and photometric time series are available in electronic format from the CDS 
via anonymous ftp to cdsarc.u-strasbg.fr (130.79.128.5) or via 
http://cdsweb.u-strabg.fr/cgi-bin/qcat?J/A+A/
}}

  \subtitle{Detected with {\it \textsc{Harps}} velocimetry. Captured in transit with {\it \textsc{Trappist}} photometry.}

\author{X.~Bonfils   \inst{1} 
   \and M.~Gillon     \inst{2}
   \and S.~Udry \inst{3}
   \and D.~Armstrong \inst{7}
   \and F.~Bouchy \inst{4}
   \and X.~Delfosse \inst{1}
   \and T.~Forveille \inst{1}
   \and E.~Jehin \inst{2}
   \and M.~Lendl \inst{3}
   \and C.~Lovis \inst{3}
   \and M.~Mayor \inst{3}
   \and J.~McCormac \inst{7}
   \and V.~Neves \inst{1,5,6}
   \and F.~Pepe \inst{3}
   \and C.~Perrier \inst{1}
   \and D.~Pollaco \inst{7}
   \and D.~Queloz \inst{3}
   \and N.~C. Santos \inst{5,6}
}

  \offprints{X. Bonfils}
	\institute{UJF-Grenoble 1 / CNRS-INSU, Institut de Plan\'etologie et d'Astrophysique de Grenoble (IPAG) UMR 5274, Grenoble, F-38041, France
	\and Institut dÕAstrophysique et de G\'eophysique, Universit\'e de Li\`ege, All\'ee du 6 Ao\^{u}t 17, Bat. B5C, 4000 Li\`ege, Belgium
          \and Observatoire de Gen\`eve, Universit\'e de Gen\`eve, 51 ch. des Maillettes, 1290 Sauverny, Switzerland
	  \and Institut d'Astrophysique de Paris, CNRS, Universit\'e Pierre et Marie Curie, 98bis Bd Arago, 75014 Paris, France
	  \and Departamento de F\'{i}sica e Astronomia, Faculdade de Ci\^encias, Universidade do Porto, Rua do Campo Alegre, 4169-007 Porto, Portugal
	  \and Centro de Astrof\'{i}sica, Universidade do Porto, Rua das Estrelas, 4150-762 Porto, Portugal
	  \and Astrophysics Research Centre, School of Mathematics \& Physics, Queen's University Belfast, University Road, Belfast BT7 1NN, UK
		}

  \date{Received  / Accepted }

 \abstract
{We report on the discovery of GJ~3470\,b, a transiting hot Uranus of mass $m_p = 14.0\pm1.8~{\rm M_\oplus}$, radius $R_p = 4.2\pm0.6~{\rm R_\oplus}$ and period $P=3.3371\pm0.0002$~day. Its host star is a nearby (d=25.2$\pm$2.9pc) M1.5 dwarf of mass $M_\star=0.54\pm0.07~{\rm M_\odot}$ and radius $R_\star=0.50\pm0.06~{\rm R_\odot}$. The detection originates from a radial-velocity campaign with \textsc{Harps} that focused on the search for short-period planets orbiting M dwarfs. Once the planet was discovered and the transit-search window narrowed to about 10\% of an orbital period, a photometric search started with \textsc{Trappist} and quickly detected the ingress of the planet. Additional observations with \textsc{Trappist}, \textit{EulerCam} and \textsc{Nites} definitely confirmed the transiting nature of GJ3470b and allow for the determination of its true mass and radius.The star's visible or infrared brightness ($V^{\rm mag}=12.3$, $K^{\rm mag}=8.0$), together with a large eclipse depth $D=0.57\pm0.05\%$, ranks GJ\,3470\,b among the most favorable planets for follow-up characterizations.}

  \keywords{stars: individual: \object{GJ~3470} --
               stars: planetary systems --
               stars: late-type --
               technique: radial-velocity, transit
              }

\titlerunning{A sub-Neptune transiting the nearby M dwarf GJ3470}
\authorrunning{X. Bonfils et al.}

  \maketitle


\section{Introduction}

Planets in transit are being detected by the thousands thanks to the $Kepler$ mission \citep{Borucki:2011}, with some being smaller than Earth \citep{Fressin:2012, Muirhead:2012} and some lying in their star's habitable zone \citep{Batalha:2012}. Kepler's detections remain however around too faint stars, and/or with too shallow transit depths, to perform transmission and occultation spectroscopy of their atmosphere. Actually, most Kepler host stars are too faint to ascertain the planetary nature of most detections or to measure their mass with current spectrographs. And therefore, observations that aim to characterize the atmosphere of exoplanets kept focused on the planets that transit bright, nearby and small stars which, so far, have been detected with ground-based instruments.

In particular, searches driven by radial-velocity (RV) observations educate the photometric observations by providing targets known to harbor planets, together with a time window to search for their possible transits. Targeting the small M dwarfs, the photometric follow-up of every planet detection is easily done from the ground with a small aperture telescope \citep[e.g.][]{Gillon:2007a, Nutzman:2008}. For instance, the transit of a Jupiter- (resp. Neptune-) like planet produces a 4-\% (resp. half-\%) drop in flux when crossing a 0.5-\Rsol star. And, since the occurrence rate of Jupiter and Neptune-like planets with short orbital periods is low \citep[below a few percent --][]{Bonfils:2011a}, their a priori RV detection and ephemeris enhance the transit-discovery power of a small telescope by a factor of a few hundred. Although smaller size planets are more frequent, their ground based detection is more difficult. One must then focus on the smallest M dwarfs \citep[e.g. ][]{Charbonneau:2009} or make use of a space-borne observatory \citep[e.g.][]{Bonfils:2011a, Demory:2011}.

This paper reports on the detection of \object{GJ 3470 b} a transiting hot Uranus detected in the framework of our radial-velocity search with \textsc{Harps} and systematic photometric follow-up. Its outline is as follows : we present the stellar properties of GJ\,3470 in Sect.~\ref{sect:prop}, the radial-velocity detection in Sect.~\ref{sect:rv} and the photometric detection in Sect.~\ref{sect:phot}. Next, we derive the posterior probability of stellar and planetary parameters for the system in Sect~\ref{sect:orb} before presenting our conclusions in Sect.~\ref{sect:concl}.

\section{\label{sect:prop}The properties of \object{GJ~3470}}
	\begin{figure}
\includegraphics[width=\linewidth]{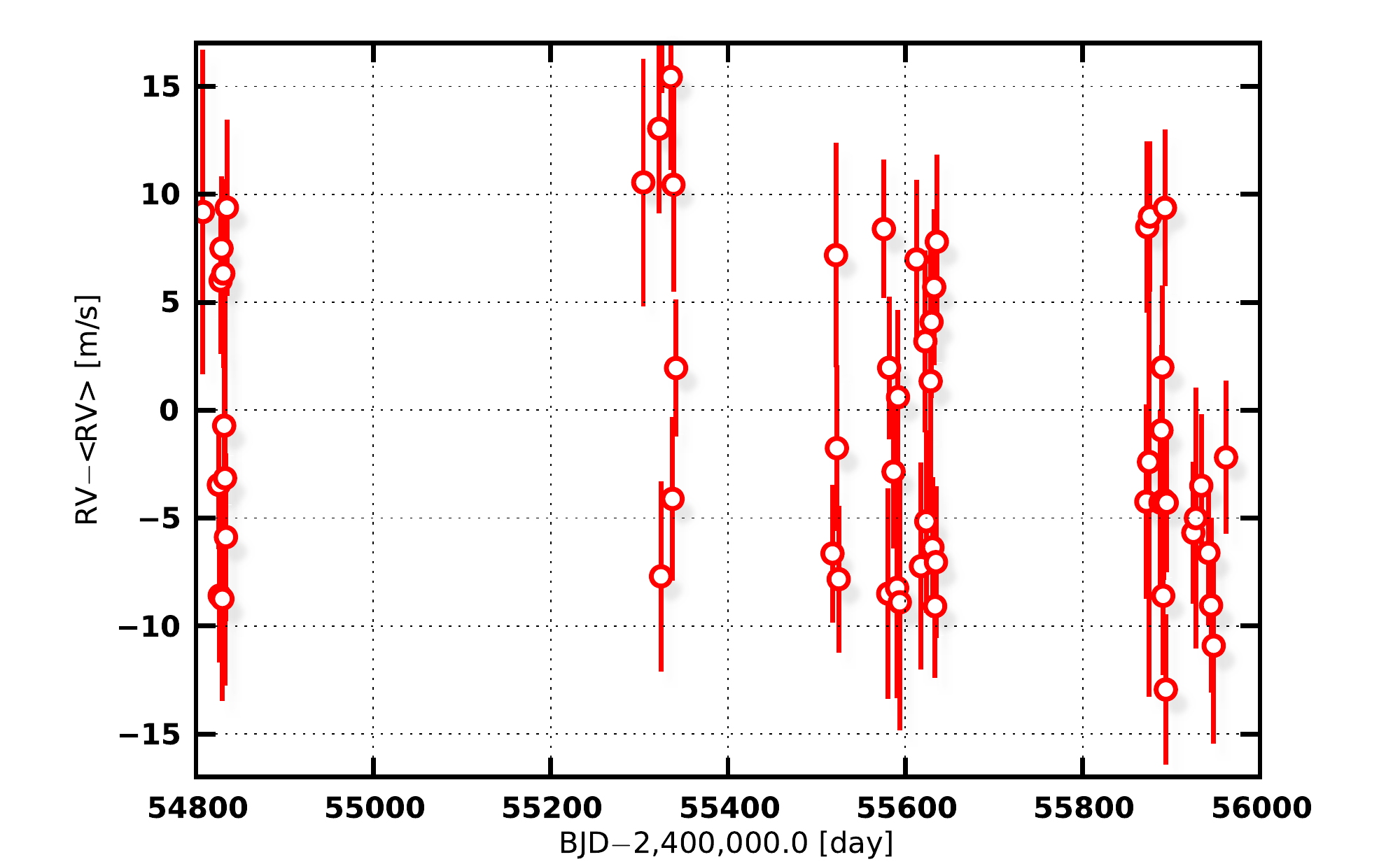}\\
\includegraphics[width=\linewidth]{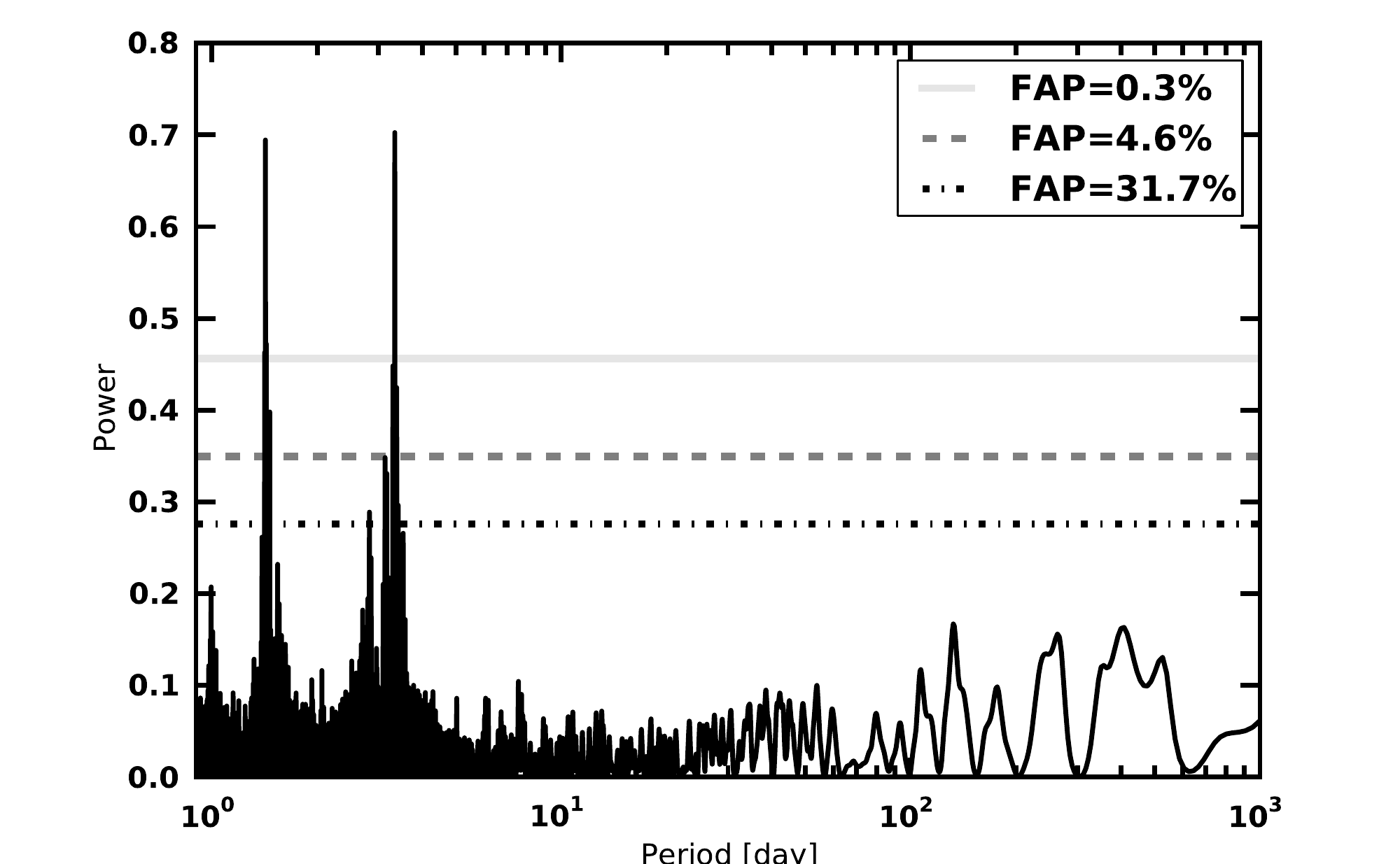}
       \caption{{\it Upper panel:}  RV time series of GJ\,3470. {\it Bottom panel:} Periodogram of GJ\,3470 RVs. The horizontal lines shows different level of false-alarm probabilities.}
       \label{fig1}
\end{figure}
\object{GJ~3470} (aka \object{LP~424-4}, \object{2MASS~J07590587+1523294}, \object{NLTT~18739}) is a M1.5 dwarf \citep{Reid:1997} seen in the Cancri constellation ($\alpha=07^h59^m06^s$, $\delta=+15^\circ23^\prime30^{\prime\prime}$) at a distance $d = 25.2\pm2.9$~pc known via photometric calibration \citep{Lepine:2005b}. 

The {\it Simbad} astronomical database also identifies \object{GJ\,3470} as \object{Melotte 25 EGG 29}, which classifies the star as a member of the Hyades cluster (Melotte 25) and could constrain its age. The denomination originates from \citet{Eggen:1990} who assigned \object{GJ\,3470} to the Hyades {\it supercluster} on the basis of its proper motion. However, Eggen also found that, to be a member of the Hyades supercluster, GJ\,3470 would have a radial velocity of about $+41.1$ km/s. For few other stars in his sample, Eggen could compare his radial velocity predictions with literature values and found good matches, most often with differences not exceeding few hundred m/s. For GJ\,3470 however, no literature radial velocity was available at the time and, today, we measure a very different value of $+26.5$ km/s. Thus, we could not consider \object{GJ\,3470} as a member of the Hyades supercluster and sought clues from other age proxies. On the one hand, we looked at GJ\,3470's $H\alpha$ (6562.808\AA) and $v \sin i$. The spectral line is seen in absorption and the projected rotational velocity is low ($\lesssim2$ km/s), both indicative of a mature star older than $\sim300$ Myr. On the other hand, we transformed GJ3470's proper motion \citep[][]{Lepine:2005a} and  systemic radial velocity (this paper) into galactic velocities (U=$+30$, V=$-12$, W=$-10$ km/s). We found they match the kinematic population of the young disk \citep{Leggett:1992}, what suggests an age of $<3$ Gyr \citep{Haywood:1997} as well as an approximately solar metallicity.

\citet{Rapaport:2001} give an apparent brightness $V^{\rm mag}=12.27\pm0.02$ in an approximately Tycho band, which is compatible with the $V^{\rm mag} \sim12.44$ estimate from \citet{Lepine:2005a}.
We used the photometric distance to convert its infrared photometry $K=7.989\pm0.023$ \citep{Cutri:2003} into an absolute magnitude $M_K^{\rm mag}=5.98\pm 0.58$ and, after bolometric  correction \citep[$BC_K^{\rm mag}=2.59$ --][]{Leggett:2001}, into a $L_\star=0.029\pm0.002$ \Lsol luminosity. The K-band mass-luminosity relation of \cite{Delfosse:2000} gives a $0.45\pm0.11 \Msol$ mass. Also, we use the theoretical mass-radius relation of \citet{Baraffe:1998}, in good agreement with interferometric measurements \citep{Demory:2009}. Assuming an age greater than 300 Myr and solar metallicity, we converted $M_\star$ into $R_\star=0.42\pm0.10$. The same models also provides mass-temperature relation and hence $T_{\rm eff} = 3600\pm200$~K. To account for limb-darkening when modeling eclipses, we used the quadratic coefficients of \citet{Claret:2000} for $T_{eff}=3500$~K, $\log g = 4.5$ and [Fe/H]=0 in the $z$-band ($u_a=0.40$, $u_b=0.19$) and V-band ($u_a=0.38$, $u_b=0.40$) filters.

\begin{table}
\caption{
\label{table:stellar}
Observed and inferred stellar parameters for GJ~3470. 
}
\begin{tabular}{l@{}lccc}
\hline
Spectral Type        &                  & M1.5                    &   R97\\
Distance, d                &[pc]           & $25.2 \pm 2.9  $    &Lep05\\
\multicolumn{4}{l}{Stellar Photometry}\\
~~~~V                            &[mag]         & $12.27\pm 0.02$& R01 \\
~~~~J                             & [mag]        & $8.794\pm0.019$ & Cu03 \\
~~~~H                           & [mag]        & $8.206\pm0.023$  & Cu03\\
~~~~K                           &  [mag]        & $7.989 \pm 0.023$&  Cu03       \\
\multicolumn{4}{l}{Stellar absolute magnitudes}\\
~~~~$M_V$                  &  [mag]        & $10.26 \pm 0.57$&\\
~~~~$M_K$                 &  [mag]        & $5.98 \pm 0.58 $&\\
Bolometric correction, $BC_K$               &  [mag]        & $2.59$ & via Leg01\\
Luminosity, $L_\star$       & [$\mathrm{L_\odot}$]    &  $0.029\pm0.002$&\\
Mass, $M_\star$       & [$\Msol$]             & $ 0.45\pm0.11 $& via D00\\
Radius, $R_\star$       & [$\Rsol$]             & $ 0.42\pm0.10 $& via B98\\
Effective temperature, T$_{\rm eff}$                 & [K]    & 3600$\pm$200 & via B98\\
Galactic velocities, (U,V,W) ~         & [km/s]                   & ($+29$, $-12$, $-10$) & \\
Age, $\tau$ & [Gyr] & $0.3$-$3$ & \\
\multicolumn{3}{l}{Limb-darkening coefficients}&Cl00\\
~~~~z' filter : $u_a$, $u_b$                &  & 0.40, 0.19 & \\
~~~~V filter : $u_a$, $u_b$               &  & 0.38, 0.40 & \\
\hline
\end{tabular}\\
Note that the stellar mass and radii are also constrained by the data of this paper and that posterior values are given in Table~\ref{table:fit}. R97: \citet{Reid:1997}; Lep05: \citet{Lepine:2005b}; R01: \citet{Rapaport:2001}; Cu03: \citet{Cutri:2003};  Leg01: \citet{Leggett:2001}; D00: \citet{Delfosse:2000}; B98: \citet{Baraffe:1998}; H97: \citet{Haywood:1997}; Cl00: \citet{Claret:2000}
\end{table}

\begin{table*}
\caption{
\label{table:fit}
Modeled and inferred parameters for the GJ~3470 system. 
}
\begin{tabular}{l@{}lcc}
\hline
                              & {\bf Unit}  & {\bf Prior} & {\bf Posterior}  \\
\hline
\multicolumn{4}{c}{Stellar parameters}\\
Stellar mass, $M_\star$       & [$\Msol$]             &  & 0.541 $\pm 0.067$\\
Stellar radius, $R_\star$       & [$\Rsol$]             &   &0.503 $\pm 0.063$\\
\hline
\multicolumn{4}{c}{Planetary parameters}\\
Orbital period, $P$                 & [day]                    & Jeffreys ($P_{\rm min}=3.33$, $P_{\rm max}=3.34$) & 3.33714 $\pm 0.00017$ \\
Systemic velocity, $\gamma$       & [km/s]                 & Uniform ($\gamma_{\rm min}=+26$, $\gamma_{\rm max}=+27$) & 26.51691 $\pm 0.00053$\\
Radial-velocity semi-amplitude, $K$ ~~~~           & [m/s]              & Jeffreys ($K_{\rm min}=0.1$,$K_{\rm max}=100$)     &0.00901 $\pm 0.00075$ \\
Orbital eccentricity, $e$                 &  [~]                          & Uniform ($e_{\rm min}=0$, $e_{\rm max}=0.1$) & $<0.051$  (1-$\sigma$ upper limit)\\
Argument of periastron, $\omega$                 &       [rad]                 & Uniform ($\omega_{\rm min}=0$, $\omega_{\rm max}=2\pi$)   & $0-2\pi$ (unconstrained) \\
Planet-to-star radius ratio, $R_p/R_\star$ &             [~]               & Uniform (${\rm min}=0$, ${\rm max}=0.1$) & 0.0755 $\pm 0.0031$\\
Transit depth, $D$ &  [mmag]                          && 5.69 $\pm 0.47$\\
Scaled semi-major axis, $a/R_\star$      &      [~]                        & Uniform (${\rm min}=0$, ${\rm max}=100$) & 14.9 $\pm 1.2$  \\
Orbital inclination, $i$                 &     [rad]                       & Isotropic spin-orbit ($i_{\rm min}=0$, $i_{\rm max}=\pi/2$)& $>88.8^{\rm o}$ (1-$\sigma$ lower limit)\\
semi-major axis, a &  [AU]                    && 0.0348 $\pm 0.0014$\\
\multicolumn{4}{l}{Transit times}\\
~~~~Mid transit, $T_{\rm tr}$       & [day]                   &&55953.6645$\pm$0.0034\\
~~~~First  contact, $t_{\rm I}$       & [day]                   &&55956.9652$\pm$0.0029\\
~~~~Second  contact, $t_{\rm II}$       & [day]                   &&55956.9705$\pm$0.0028\\
~~~~Third  contact, $t_{\rm III}$       & [day]                   &&55957.0325$\pm$0.0051\\
~~~~Fourth  contact, $t_{\rm IV}$       & [day]                   &&55957.0325$\pm$0.0051\\
Planetary mass, M$_p$                 & [M$_\oplus$]    && 14.0 $\pm 1.7$ \\
Planetary radius, R$_p$                 & [R$_\oplus$]    &&4.2 $\pm 0.6$  \\
Impact parameter, $b$                  &     [~]                          &&0.22 $\pm 0.16$ \\
Planetary density, $\rho_{p}$                 & [g/cm$^3$]    &&1.07 $\pm 0.43$\\
Planetary surface gravity, g$_p$                & [m/s$^2$]    &&7.9 $\pm 1.8$ \\
\multicolumn{4}{l}{Equilibrium temperature}\\
~~~~T$_{eq}$, A=0 &  [K]                    &&615 $\pm 16$\\
~~~~T$_{eq}$, A=0.75 &  [K]                  &  &435 $\pm 12$\\
\hline
\multicolumn{4}{c}{Data parameters}\\
\multicolumn{4}{l}{Radial-velocity time series additive noise}\\
~~~~for {\it Harps}, $\epsilon_{\rm rv}$ ~& [m/s] &Modif. Jeffreys  ($\epsilon_{\rm min}=0$,$\epsilon_{\rm 0}=0.1$,$\epsilon_{\rm max}=1000$)& $<0.40$  (1-$\sigma$ upper limit)\\
\multicolumn{4}{l}{Photometric time series additive noise}\\
~~~~for {\it \textsc{Trappist}} (2012, Feb. 26th), $\epsilon_{\rm ph,1}$ ~& [mmag] &Modif. Jeffreys  ($\epsilon_{\rm min}=0$, $\epsilon_{\rm 0}=0.1$, $\epsilon_{\rm min}=1000$)& 2.18 $\pm 0.19$\\
~~~~for {\it \textsc{Trappist}} (2012, Mar.   7th), $\epsilon_{\rm ph,2}$ & [mmag] &Modif. Jeffreys ($\epsilon_{\rm min}=0$, $\epsilon_{\rm 0}=0.1$, $\epsilon_{\rm min}=1000$)&3.51 $\pm 0.18$ \\
~~~~for {\it Euler}      (2012, Feb. 26th), $\epsilon_{\rm ph,3}$ & [mmag] &Modif. Jeffreys ($\epsilon_{\rm min}=0$, $\epsilon_{\rm 0}=0.1$, $\epsilon_{\rm min}=1000$)&1.33 $\pm 0.12$ \\
~~~~for {\it \textsc{Nites}}      (2012, Apr. 12th), $\epsilon_{\rm ph,4}$ & [mmag] &Modif. Jeffreys ($\epsilon_{\rm min}=0$, $\epsilon_{\rm 0}=0.1$, $\epsilon_{\rm min}=1000$)&0.04 $\pm 0.04$ \\
\multicolumn{4}{l}{Baseline model ($y=a +b \tau+c \tau^2$, where $\tau$ is, for a given time series, the time elapsed since the 1st exposure)}\\
~~~~for {\it \textsc{Trappist}} (2012, Feb. 26th), $a$ & & & $1.000100\pm0.000461$ \\
~~~~~~~~~~~~~~~~~~~~~~~~~~~~~~~~~~~~~~~~~~~~~~~~~~~~, $b$&&& $-0.011\pm0.012$ \\
~~~~~~~~~~~~~~~~~~~~~~~~~~~~~~~~~~~~~~~~~~~~~~~~~~~~, $c$&&& $0.097\pm0.068$ \\
~~~~for {\it \textsc{Trappist}} (2012, Mar.   7th), $a$ & & & $0.999810\pm0.000502$ \\
~~~~~~~~~~~~~~~~~~~~~~~~~~~~~~~~~~~~~~~~~~~~~~~~~~~, $b$&&& $-0.003\pm0.011$ \\
~~~~~~~~~~~~~~~~~~~~~~~~~~~~~~~~~~~~~~~~~~~~~~~~~~~, $c$&&& $0.049\pm0.053$ \\
~~~~for {\it Euler}      (2012, Feb. 26th), $a$ & & & $0.998510\pm0.000355$ \\
~~~~~~~~~~~~~~~~~~~~~~~~~~~~~~~~~~~~~~~~~~~~~~~, $b$&&& $0.068\pm0.010$ \\
~~~~~~~~~~~~~~~~~~~~~~~~~~~~~~~~~~~~~~~~~~~~~~~, $c$&&& $-0.487\pm0.061$ \\
~~~~for {\it \textsc{Nites}}      (2012, Apr. 12th), $a$ & & & $0.186031\pm0.000083$ \\
~~~~~~~~~~~~~~~~~~~~~~~~~~~~~~~~~~~~~~~~~~~~~~~, $b$&&& $-0.0058\pm0.0027$ \\
~~~~~~~~~~~~~~~~~~~~~~~~~~~~~~~~~~~~~~~~~~~~~~~, $c$&&& $0.302\pm0.018$ \\
\hline
\end{tabular}\\
\end{table*}

\section{\label{sect:rv}Spectroscopic detection}

We observed \object{GJ\,3470} with the \textsc{Harps} spectrograph, the state-of-the-art velocimeter fiber-fed by the ESO/3.6-m telescope \citep{Mayor:2003, Pepe:2004}. Our settings remained the same as for previous observations \citep[e.g.][]{Bonfils:2011a}. Like \object{GJ3634} --that hosts a super-Earth reported by our survey \citep{Bonfils:2011b}--, \object{GJ\,3470} is part of an extended sample of $\sim$300 M dwarfs specifically targeted to search for short-period planets, and the subset that transit. 

We collected 61 radial-velocity observations of GJ\,3470 between 2008 Dec 08 and 2012 Jan 14. All are 900-s exposures except the first one that is a 300-s exposure used to verify the star is suitable for a planet search (i.e. not too active, not too fast a rotator, nor a spectroscopic binary). Their median uncertainty is $\sigma_i=3.7$~m/s and result predominantly from photon noise. We note this uncertainty agrees well with that of other M dwarfs with similar brightness. 
The GJ\,3470 time series (shown in the top panel of Fig.~\ref{fig1} and available electronically) has an observed dispersion $\sigma_e=7.7$~m/s, indicating intrinsic variability. The periodicity of that signal is seen in a generalized Lomb-Scargle periodogram \citep[Fig.~\ref{fig1}, bottom panel --][]{Press:1992, Zechmeister:2009b} where prominent power excesses are exhibited at periods $P\sim3.33$ and 1.42 day (alias of each other for a 1-day time sampling). The periodogram is shown in Fig.~\ref{fig1}, together with false-alarm probability levels computed with bootstrap randomization of the original data \citep[see][for more detailed descriptions]{Bonfils:2011a, Bonfils:2011b}. 

To model the data with a Keplerian orbit we chose a Bayesian framework and employed a Markov Chain Monte Carlo algorithm \citep[MCMC -- e.g.][]{Gregory:2005, Gregory:2007, Ford:2005}. MCMC algorithms sample the joint probability distribution for the model parameters by evolving a solution (i.e. a set of parameter values) in the manner of a random walk. At each step, a new solution is proposed to replace the previous solution. The new solution is accepted following a pseudo-random process that depends on the $\chi^2$ difference between both solutions such that solutions with a higher likelihood are accepted more often. Step-by-step, accepted solutions build a {\it chain} which, after enough iterations, reaches a stationary state. One can then discard the first iterations and keep only the stationary part of the chain. The distributions of parameter values of all the remaining chain links then correspond to the targeted  joint probability distribution for the model parameters. Our implementation follows closely the one of \citet{Gregory:2007} with several (10 in our case) chains running in parallel. Each chain is attributed a parameter $\beta$ that scales the likelihood such that chains with a lower $\beta$ value present a higher acceptance probability. We also paused the MCMC iteration after each 10 steps and proposed the chains to permute their solutions (which was again accepted pseudo-randomly and according to the likelihood difference between solutions). This approach is reminiscent of simulated  annealing algorithms and permit evasion outside of local minima and better exploration of the wide parameter space. Only the chain with $\beta=1$ corresponds to the targeted probability distribution. Eventually, we thus discarded all chains but the one with $\beta=1$. We adopted the median of the posterior distributions for the optimal parameter values, and the 68\% centered interval for their uncertainties. Hence, the orbital parameters were notably $P=3.3368\pm0.0004$ day, $K=9.2\pm0.8$ m/s, $e=0.16\pm0.08$ and inferior conjunction $T_{\rm tr} = 2455983.42\pm0.14$~ JD, modulo integer multiples of the period $P$. Solutions with period $P\sim1.42$ day were also found, albeit with a 10 times lower occurence. 
 
For a $M_\star=0.45\pm0.11 \Msol$ star, the optimal parameter values correspond to a companion with a mass $m_p \sin i = 12\pm2 \Mearth$.

\section{\label{sect:phot}Photometric detection}
From the radial velocity orbit, we could predict hypothetic transit times with a 0.14-day accuracy. From mass-radius relationships \citep[e.g.][]{Fortney:2007}, we also estimated the transit depth caused by the eclipse of a pure iron planet would be of the order of 1 milli-magnitude and measurable from the ground. We aimed to perform a photometric search for a $\pm2$-$\sigma$ window, i.e. 95\% of the posterior density function (PDF) of transit times.
	\begin{figure*}
\includegraphics[width=.5\linewidth]{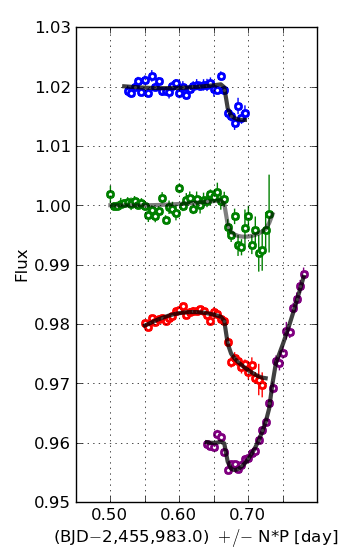}
\includegraphics[width=.5\linewidth]{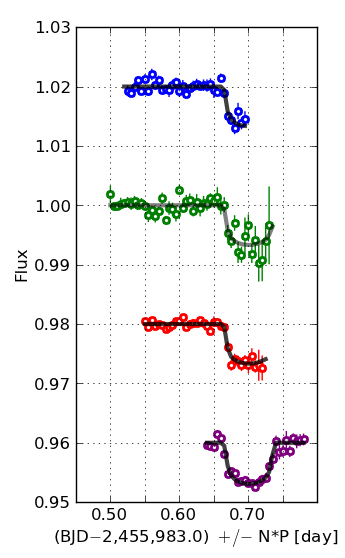}
       \caption{Light curves of GJ\,3470. Left panel corresponds to raw light curves (with arbitrary offsets for clarity). Right panel corresponds to de-trended light curves. From top to bottom, light curves were obtained with \textsc{Trappist} (on 2012 February 26th), \textsc{Trappist} (on 2012 March 7th), {\it EulerCam} (on 2012 March 7th) and \textsc{Nites} (on 2012 April 12th). The optimal model is over-plotted with a black curve (see Sect.~\ref{sect:orb}).}
       \label{fig2}
\end{figure*}

\subsection{First transit ingress detection with \textsc{Trappist}}
 We initiated the photometric search with \textsc{Trappist} and observed on 2012, February 26th between 00$^{\rm h}$25$^{\rm m}$ and 04$^{\rm h}$36$^{\rm m}$ UT. \textsc{Trappist} is a robotic 60-cm telescope installed at La Silla Observatory. It is equipped with a 2048 x 2048 15$\mu$m CCD camera, provides a 22'x22' filed-of-view with a 0.65'' pixel scale \citep{Jehin:2011}. We used a z'-Gunn filter, made 764 exposures of 10 s each and recorded flux for GJ\,3470 together with carefully chosen comparison stars in the field. We applied the reduction procedure described in \citet{Gillon:2011} to derive the differential photometry (which time series are available in ascii form electronically). We note that changing the comparison stars did not affect the results of our reduction. In Fig.~\ref{fig2} (top) we show the time series after applying binning by 0.005 day (7.2 min) and after computing uncertainties by measuring the dispersion in each bin and divided by the square-root of the number of points.
 
Right after the first night of follow-up we identified a drop in flux compatible with the ingress of a $\sim4$-R$_\oplus$ planet, seen in Fig.~\ref{fig2} to start at BJD=2,455,983.66 day. The transit seems to occur somewhat less than 2$\sigma$ later than predicted. We inspected Digital Sky Survey (DSS) images to look for a possible background star. The proper motion of GJ\,3470 is fast and for its present position, the 60-year old DSS images can confidently exclude stars more than 5 mag fainter than GJ3470. We noted the event is not compatible with the aliased 1.4-d period and we pursued the follow-up with \textsc{Trappist}, the {\it Euler-Swiss} and {\it Nites} telescopes with adjusted ephemeris.

\subsection{Confirmation with {\it EulerCam}, \textsc{Trappist} and \textsc{Nites}}
From La Silla Observatory, and after our first (fortunate) night of follow-up, only one transit event could be visible before the end of the current season. The ingress was foreseen at BJD$\simeq$2,455,993.68 (2012 March 7th, 04$^{\rm h}$19$^{\rm m}$ UT) at an airmass greater than 1.8. To secure the confirmation we made use of {\it EulerCam} in complement of \textsc{Trappist}. {\it EulerCam} is a 4k$\times$4k CCD camera mounted on the Euler Swiss telescope that is also installed in La Silla Observatory. The field-of-view of {\it EulerCam} is somewhat smaller than the one of \textsc{Trappist} (15.7'$\times$15.7') but the twice as large telescope aperture mitigates the scintillation by a factor $\sim1.5$. 

\textsc{Trappist} started observing at BJD$=$2,455,993.509 (00$^{\rm h}$13$^{\rm m}$ UT), recorded 943 10-s exposures, and stopped at BJD$=$2,455,993.743 (05$^{\rm h}$50$^{\rm m}$ UT) when GJ\,3470 reached an airmass $z=3.55$. {\it EulerCam} started at BJD$=$2,455,993.559 (01$^{\rm h}$25$^{\rm m}$ UT), recorded 223 50-s exposures, and stopped at BJD$=$2,455,993.733 (05$^{\rm h}$36$^{\rm m}$ UT) when GJ\,3470 reached an airmass $z=3.04$. We used z'-Gunn filter with both telescopes. The reduction procedure used for the {\it EulerCam} is described in \citet{Lendl:2012}. The time-series photometry presented has been obtained with relative aperture photometry, where apertures and references were chosen carefully. As for the previous light-curve, both time-series were binned by 0.005 day and shown in Fig.~\ref{fig2} (second and third curves, from top to bottom).

On both light curves, we identified a 6.6$\pm0.4$ mmag drop in flux consistent with the transit of a $\sim$4-$R_\oplus$ planet, with an ingress timing in sync with our predictions. Unfortunately, the star reached a too high airmass for photometric observations before we could record the transit egress.

To record a full transit event we continued the follow-up in the Northern hemisphere, at La Palma, where we predicted the visibility of few more events before the star would go behind the Sun. The La Palma observations were obtained with the 0.4m \textsc{Nites} telescope \citep{McCormac:2012}. Images were obtained in white light using a 1024$\times$1024
deep depleted CCD giving 0.7"/pixel and a 12'$\times$12' field of
view. The entire transit was observed without filter and at high airmass. The data were
reduced with standard IRAF routines.
Observations started on BJD=2,460,030.357 (2012 April 12th, 20$^{\rm h}$34$^{\rm m}$ UT) and stopped on BJD=2,460,030.500 (2012 Apr 13th, 00$^{\rm h}$00$^{\rm m}$ UT). We recorded 586 exposures of 20 s which aperture photometry is also shown at the bottom of Fig.~\ref{fig2}, after binning by 0.005 day. The light curve appears with a trend of large amplitude which is most probably caused by the color-difference between our target and the comparison stars. The airmass-extinction depends on the stars' colors and our target is sensibly redder than the average comparison star in the field, an effect that is difficult to mitigate for unfiltered observations. Nevertheless the light curve does confirm the transit event and, more importantly, does constrain its duration.

	\begin{figure}
\includegraphics[width=\linewidth]{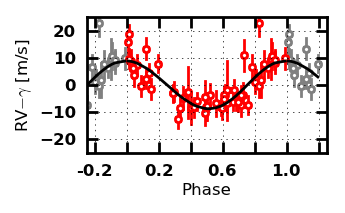}\\
\includegraphics[width=\linewidth]{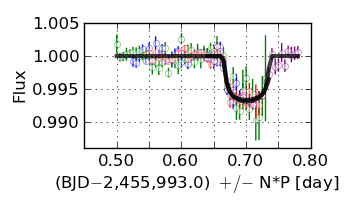}\\
       \caption{RV (top) and photometry time series (bottom) phased to the planet's orbital period and with the optimal model over-plotted. A 2-color code (gray and red) is used for RVs to mark points shown twice. The photometry color code is the same as for Fig.~\ref{fig2}.}
       \label{fig3}
\end{figure}

\section{\label{sect:orb}Joined RV$+$photometry modeling}

	\begin{figure}
\includegraphics[width=\linewidth]{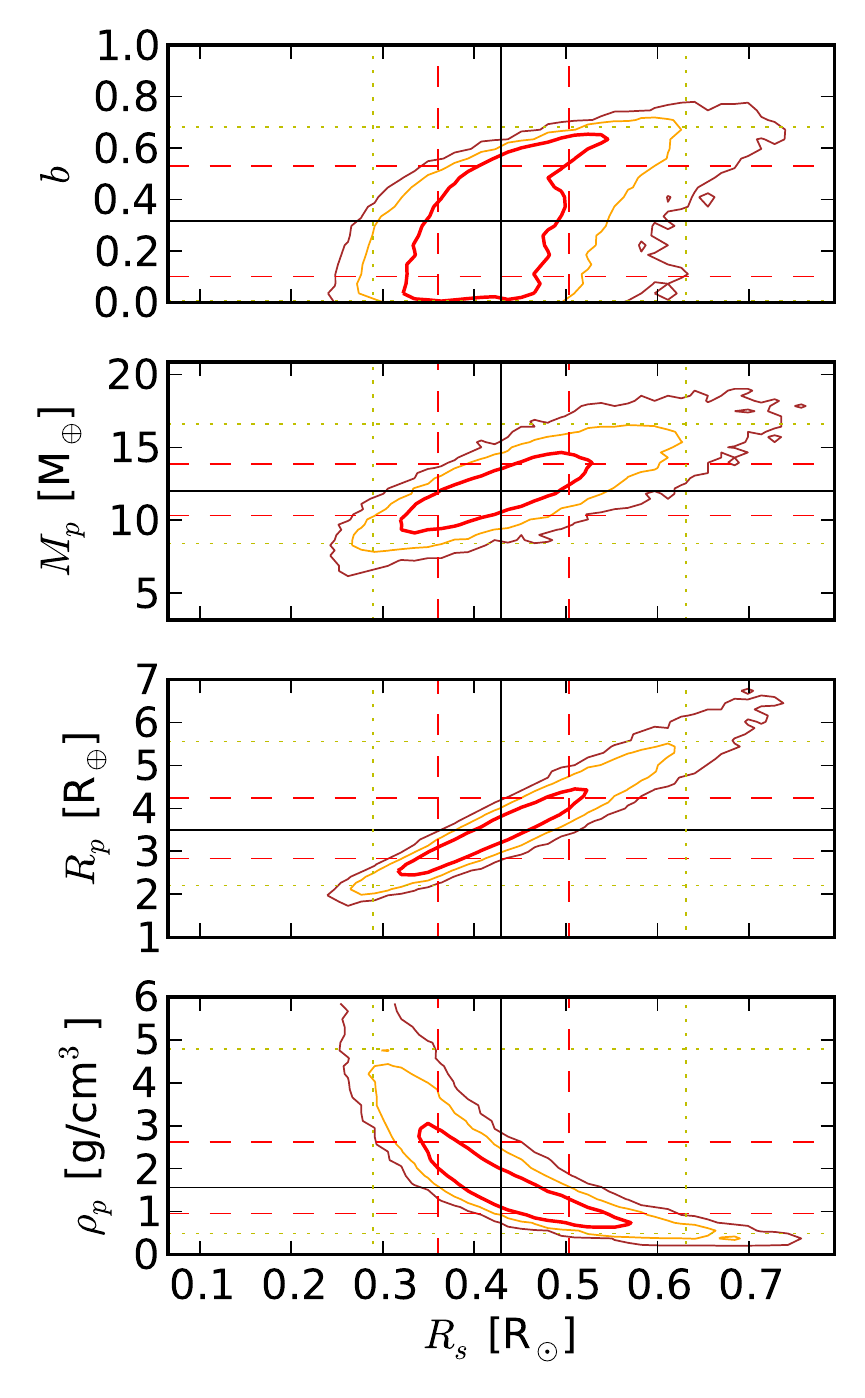}
       \caption{\label{fig:bR}Posterior distributions for selected system parameters. Iso-contours delineate 68.2, 95.4 and 99.7\% confidence intervals (red, orange and brown, respectively). The median, 68\% and 99.7\% confidence intervals of marginalized single parameters are also reported with continuous-black, dash-red and dot-yellow lines, respectively}\end{figure}

To measure physical and orbital parameters of the GJ\,3470 planetary system, we pooled together the photometry and radial-velocity time series. As for the spectroscopic orbit (Sect.~\ref{sect:rv}), we chose a Bayesian framework and used a MCMC algorithm. We modeled the data with a planet on a Keplerian motion around the star, with 18 variables for the parametrization : the systemic velocity $\gamma$, the orbital period $P$, the RV semi-amplitude $K$, the eccentricity $e$, the argument of periastron $\omega$, the time of passage at periastron $T_0$, the planet-to-star radius ratio $R_p/R_\star$, the scaled semi-major axis $a/R_\star$ and the orbital inclination on the sky $i$. In addition, to allow for quadratic baselines in the photometric time-series we add 4$\times$3 parameters, and to model the additive error (i.e. quadratically in excess of photon noise) we used 5 ad hoc variables, $\epsilon_{\rm rv}$, $\epsilon_{\rm ph, 1}$, $\epsilon_{\rm ph, 2}$, $\epsilon_{\rm ph, 3}$ and $\epsilon_{\rm ph, 4}$, for each one of the 5 RV and photometric time series.

\begin{figure*}
\centering
\includegraphics[width=\linewidth]{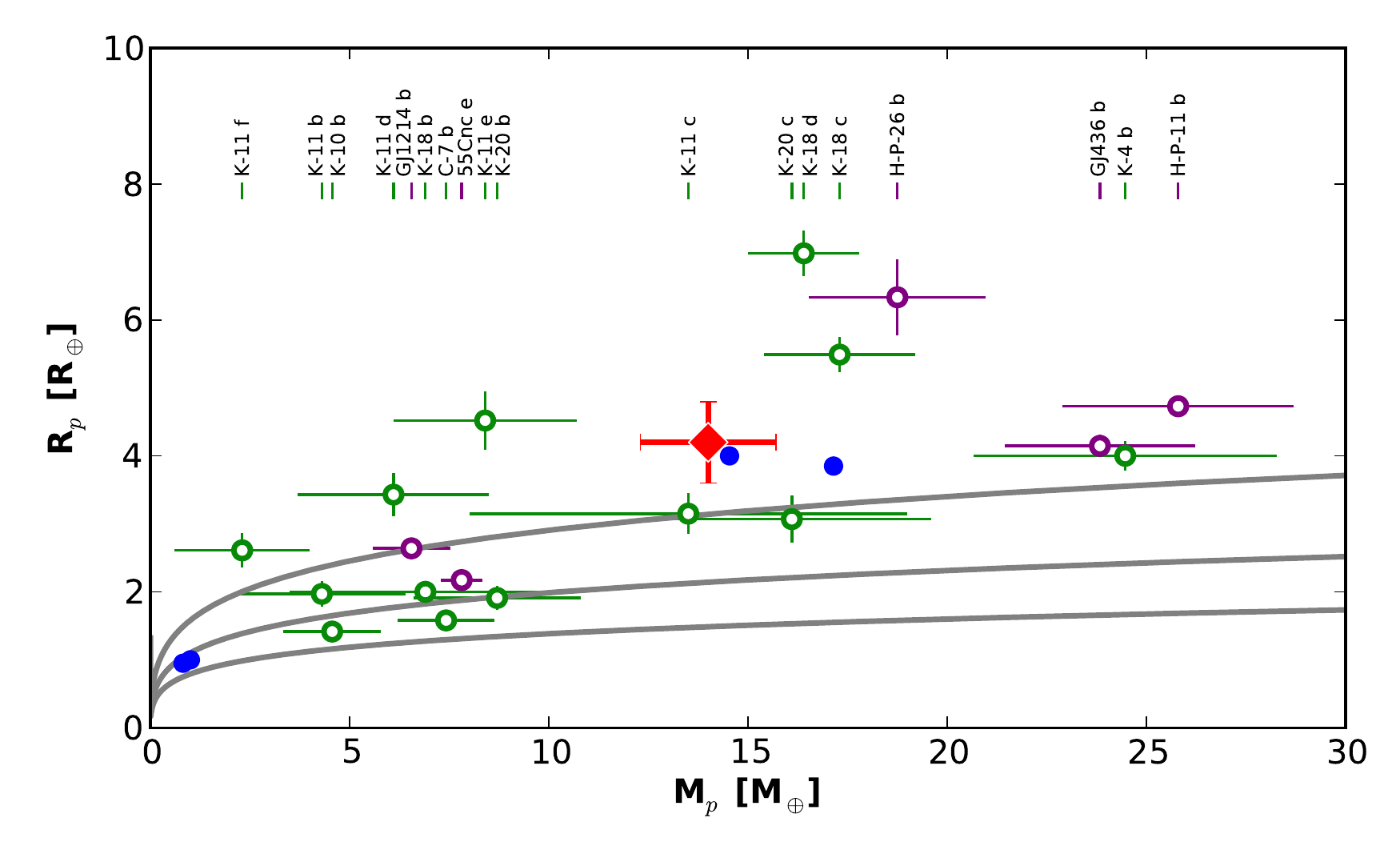}
       \caption{\label{fig:MR} Diagram with measured masses and radii for known exoplanets. Blue empty circles are for Venus, Earth, Uranus and Neptune (from left to right). Purple points are for previously known transiting exoplanets detected from the ground and green points are for planets detected from space by Kepler or CoRoT. Small vertical ticks on the top of the figure help label each detection. Kepler, CoRoT and HAT planets have short labels, K, C and H, respectively. GJ\,3470\,b is over plotted with a red diamond. The gray curves show mass-radius relations for water, rock (MgSiO$_3$ perovskite) and iron, from top to bottom, respectively \citep{Fortney:2007}.
       }
\end{figure*}
As dictated by Bayesian statistics we gave {\it a priori} descriptions for the probability distributions of parameter values. We chose uniform probability distributions for $\gamma$, $e$, $\omega$, $T_{0}$, $a/R_\star$, $R_p/R_\star$, $\log(P)$ and $\log(K)$ (so called {\it Jeffreys} priors for $P$ and $K$). We chose $i$ such that orbits have isotropic distributions. And, we chose distributions that are uniform in a logarithmic scale above a given threshold and uniform in a linear scale below (so called {\it modified Jeffreys} priors), for all 4 additive noise parameters. With each prior distribution we defined a range of authorized values (labeled by $min$ and $max$ subscripts) and,  a threshold value for the modified Jeffreys priors (labeled with $0$ subscripts). 
Only for the baselines parameters, that are all linear, we solved for the best $\chi^2$ analytically. We report our choices for the prior distributions, parameters ranges and threshold values in Table~\ref{table:fit}. 

In addition, we assumed a stellar mass-radius relationship by interpolating within the \citet{Baraffe:1998} models grid (with [M/H]=0~dex, Y=0.275, Lmix=1.0 and Age=1~Gyr), and considered the stellar mass determination from Sect.~\ref{sect:prop} as one more piece of observational data rather than a parameter to fit. Note that we chose not to include a description of the Rossiter-McLaughlin effect. It is thought to be small compared to the photon noise and, more practically, no radial-velocity happened to be taken during a transit event.

To enhance the convergence efficiency, we restricted the intervals explored for the orbital period. This allowed us to converge without using several chains of different temperatures running in parallel. We checked that the interval we chose was much larger than the posterior distributions for orbital periods and, therefore, did not affect our results.   
The MCMC chain converged on a stationary solution after $\sim$10,000 steps. We continued iterations for 500,000 more steps and inferred the posterior distributions of parameters from those last steps only. We adopted the median of the posterior distributions for the optimal parameter values, and the 68\% centered interval for their (``1 $\sigma$'') uncertainties (Table~\ref{table:fit}).

We also picture the optimal model emerging from our stochastic fitting with Fig.~\ref{fig3}, where it is plotted over the phase-folded RV and photometry data. Fig.~\ref{fig:bR} moreover shows the posterior distributions for a subset of system parameters. The top panel of that figure shows that our estimate for the planet's impact parameter is correlated to the stellar radius. For a small star, only a central transit matches the duration of La Palma's event. A larger star allows for a wider range of impact parameters but too large a star is rendered improbable by our prior input on $M_\star$. 
Our global modeling also attributes a mass, radius and thus density to the planet, $M_p = 14.0\pm1.7~{\rm M}_\oplus$, $R_p =4.2\pm0.6 {\rm R}_\oplus$ and $\rho_p=1.07\pm0.43$ g/cm$^3$, respectively. 

We repeated the above analysis with polynomials of higher degrees for the photometric baselines. NITES photometric trend is the strongest and we thus paid a particular attention to the egress timings. We tested 2nd, 3rd and 4th order polynomials and found slight changes in the timings determinations (consistent with the $1-\sigma$ uncertainties quoted in Table 2). We also found the stellar radius determination and correlated parameters $M_\star$, $M_p$ and $R_p$ are sensitive to the choice of baseline. With higher degree polynomials, smaller values are found for all parameters (e.g. $M_\star=0.46\pm0.08 {\rm M_\odot}$, $R_\star=0.43\pm0.07 {\rm R_\odot}$, $M_p=12.0\pm1.8 {\rm M_\oplus}$ and $R_p=3.49\pm0.72 {\rm R_\oplus}$ for the 4th-order polynomial). 

Both Fig.~\ref{fig:bR} and the susceptibility to baseline choices show well that most of the uncertainties for those planetary parameters are bounded to the large uncertainties on the stellar properties (radius or mass). Fortunately, the possibility to refine both stellar and planetary parameters with more precise light curves is a virtue of planetary transits \citep{Seager:2003}. For instance, a light curve of higher quality, such as those produced with $\it Spitzer$, shall refine ingress and egress durations. This will result in a strong constraint on the impact parameter, resolve the $b$-$R_\star$ degeneracy and improve the precision of the planetary mass, radius and density.

\section{\label{sect:concl}Conclusion}

We have presented the transiting planet GJ\,3470\,b discovered with \textsc{Harps} radial velocities and subsequent photometric follow-up with \textsc{Trappist}, the {\it Euler-Swiss} and the \textsc{Nites} telescopes. The planet detection adds to the small subset of low-mass planets ($M_p\lesssim 30 \Mearth$) with measured masses and radii (Fig.~\ref{fig:MR}). Today, its bulk properties remain largely unconstrained, mostly because we have limited knowledge on the stellar properties. With the current mass and radii determinations, GJ\,3470\,b would be an ice giant and seems comparable to Uranus in our Solar System. Among transiting exoplanets, it is intermediate to, on the one hand GJ\,436\,b \citep{Butler:2004, Gillon:2007a, Gillon:2007b}, HAT-P-26\,b \citep{Hartman:2011c} and HAT-P-11\,b \citep{Bakos:2010} and, on the other hand, GJ\,1214 \citep{Charbonneau:2009} and 55 Cnc e \citep{Winn:2011, Demory:2011, Gillon:2012}. It is actually close to the nominal mass and radius of Kepler-11\,c \citep{Lissauer:2011a} and Kepler-20\,c \citep{Gautier:2012} which have loose mass determinations.

Resolving the parameter degeneracy is likely to place GJ\,3470\,b as a remarkable planet. Indeed, on the one hand, if future observations attribute a small radius (e.g. $R_\star \sim 0.45 \Rsol$) to GJ\,3470, the planet will be of low-mass ($M_p\sim12\Mearth$) and small ($R_p\sim3.8\Rearth$), in a yet unpopulated mass-radius domain. On the other hand, if future observations confirm or inflate the moderate radius of GJ\,3470 ($R_\star \gtrsim 0.50 \Rsol$), the planet will have an unusual low density ($\rho_p\lesssim1.07$ g/cm$^3$) compared to other ice giants (e.g. $\rho=1.3$ and 1.6 g/cm$^3$ for Uranus and Neptune, respectively).

\begin{figure}
\centering
\includegraphics[width=\linewidth]{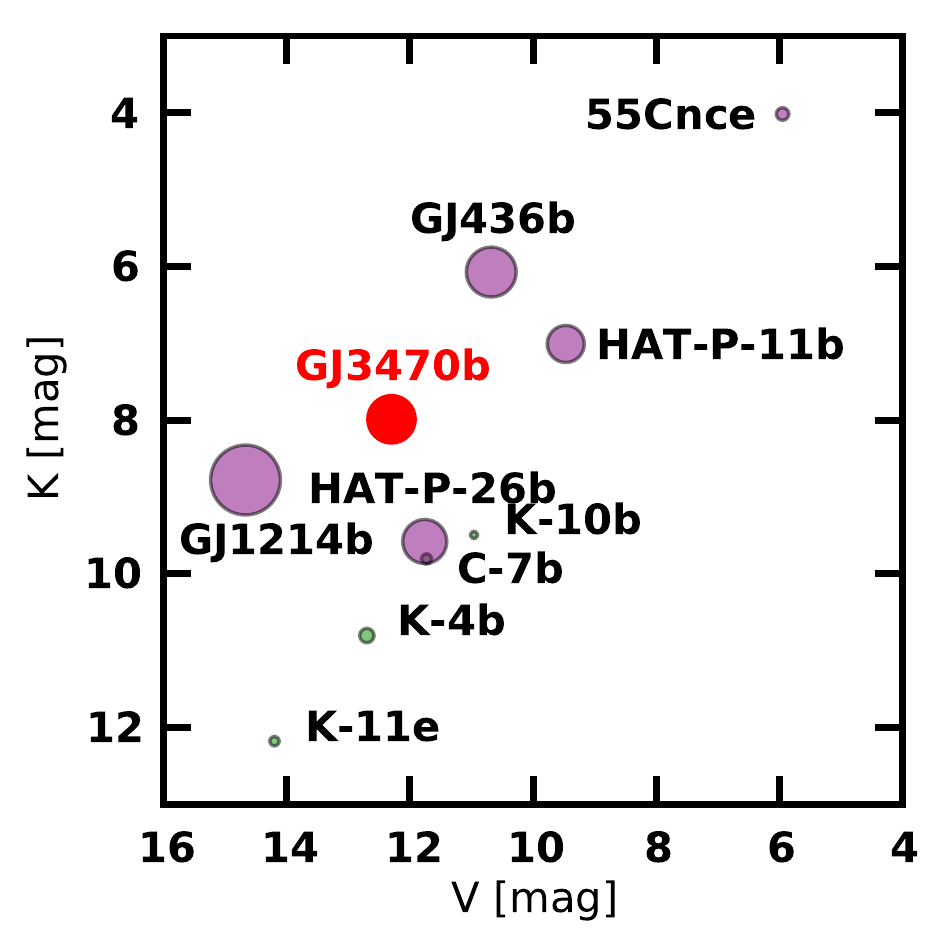}
       \caption{\label{fig:kvd} Attributes for follow-up suitability : K- and V-band stellar brightness for the coordinates and transit depth for the symbol size. Kepler and CoRoT planets are abbreviated with K and C. Adapted from \citet{Henry:2011}.  
       }
\end{figure}

Perhaps more importantly, the GJ\,3470 system has favorable attributes for follow-up characterization (V$^{\rm mag}$=12.3, K$^{\rm mag}$=7.99, D=0.62\%). Among super-Earths and exo-Neptunes only 55 Cnc\,e, GJ\,436\,b and HAT-P-11\,b transit stars brighter in both V- and K-bands. Among those 3, only GJ\,436\,b produces a larger transit depth (see Fig.~\ref{fig:kvd}). 

We thus encourage follow-up measurements of GJ\,3470\,b. Improved transit light curves should increase the precision of the planetary and stellar parameters. Chromatic light-curves \citep[e.g.][]{Bean:2011, Berta:2012} could probe the physico-chemistry of GJ\,3470\,b's atmosphere. On longer time-scales light curves could also measure GJ\,3470's rotation and constrain its age with gyrochronology relations \citep{Barnes:2007, Delorme:2011}. Astrometry could give a trigonometric parallax and improve knowledge of GJ\,3470's distance. Finally, as the orbits of multi-planet systems are thought to have low mutual inclinations \citep{Lissauer:2011b, Figueira:2012}, the low value of planet b's impact parameter ($<0.17\pm0.13$) makes of GJ\,3470 an attractive target to search for additional transiting/cooler planets. 

\begin{acknowledgements}
\end{acknowledgements}
We wish to thank David Ehrenreich for fruitful discussions. We are grateful to ESO La Silla staff for its continuous support. TRAPPIST is a project funded by the Belgian Fund for Scientific Research (Fond National de la Recherche Scientifique, F.R.S-FNRS) under grant FRFC 2.5.594.09.F, with the participation of the Swiss National Science Fundation (SNF).  M. Gillon and E. Jehin are FNRS Research Associates. NCS and VN acknowledge the support by the European Research Council/European Community under the FP7 through Starting Grant agreement number 239953. NCS also acknowledges the support from Funda\c{c}\~ao para a Ci\^encia e a Tecnologia (FCT) through program Ci\^encia\,2007 funded by FCT/MCTES (Portugal) and POPH/FSE (EC), and in the form of grants reference PTDC/CTE-AST/098528/2008 and PTDC/CTE-AST/098604/2008. VN would like to acknowledge the support from the FCT in the form of the fellowship SFRH/BD/60688/2009.

\bibliographystyle{aa}
\bibliography{mybib}

\begin{thebibliography}{51}
\expandafter\ifx\csname natexlab\endcsname\relax\def\natexlab#1{#1}\fi

\bibitem[{Bakos {et~al.}(2010)Bakos, Torres, P{\'a}l, Hartman, Kov{\'a}cs,
  Noyes, Latham, Sasselov, Sip{\H o}cz, Esquerdo, Fischer, Johnson, Marcy,
  Butler, Isaacson, Howard, Vogt, Kov{\'a}cs, Fernandez, Mo{\'o}r, Stefanik,
  L{\'a}z{\'a}r, Papp, \& S{\'a}ri}]{Bakos:2010}
Bakos, G.~{\'A}., Torres, G., P{\'a}l, A., {et~al.} 2010, \apj, 710, 1724

\bibitem[{Baraffe {et~al.}(1998)Baraffe, Chabrier, Allard, \&
  Hauschildt}]{Baraffe:1998}
Baraffe, I., Chabrier, G., Allard, F., \& Hauschildt, P.~H. 1998, \aap, 337,
  403

\bibitem[{Barnes(2007)}]{Barnes:2007}
Barnes, S.~A. 2007, \apj, 669, 1167

\bibitem[{Batalha {et~al.}(2012)Batalha, Rowe, Bryson, Barclay, Burke,
  Caldwell, Christiansen, Mullally, Thompson, Brown, Dupree, Fabrycky, Ford,
  Fortney, Gilliland, Isaacson, Latham, Marcy, Quinn, Ragozzine, Shporer,
  Borucki, Ciardi, Gautier, Haas, Jenkins, Koch, Lissauer, Rapin, Basri, Boss,
  Buchhave, Charbonneau, Christensen-Dalsgaard, Clarke, Cochran, Demory,
  DeVore, Esquerdo, Everett, Fressin, Geary, Girouard, Gould, Hall, Holman,
  Howard, Howell, Ibrahim, Kinemuchi, Kjeldsen, Klaus, Li, Lucas, Morris, Prsa,
  Quintana, Sanderfer, Sasselov, Seader, Smith, Steffen, Still, Stumpe, Tarter,
  Tenenbaum, Torres, Twicken, Uddin, Cleve, Walkowicz, \& Welsh}]{Batalha:2012}
Batalha, N.~M., Rowe, J.~F., Bryson, S.~T., {et~al.} 2012, eprint arXiv, 1202,
  5852

\bibitem[{Bean {et~al.}(2011)Bean, D{\'e}sert, Kabath, Stalder, Seager,
  Kempton, Berta, Homeier, Walsh, \& Seifahrt}]{Bean:2011}
Bean, J.~L., D{\'e}sert, J.-M., Kabath, P., {et~al.} 2011, eprint arXiv, 1109,
  582, submitted to ApJ

\bibitem[{Berta {et~al.}(2012)Berta, Charbonneau, D{\'e}sert, Kempton,
  McCullough, Burke, Fortney, Irwin, Nutzman, \& Homeier}]{Berta:2012}
Berta, Z.~K., Charbonneau, D., D{\'e}sert, J.-M., {et~al.} 2012, \apj, 747, 35

\bibitem[{Bonfils {et~al.}(2011{\natexlab{a}})Bonfils, Delfosse, Udry,
  Forveille, Mayor, Perrier, Bouchy, Gillon, Lovis, Pepe, Queloz, Santos,
  S{\'e}gransan, \& Bertaux}]{Bonfils:2011b}
Bonfils, X., Delfosse, X., Udry, S., {et~al.} 2011{\natexlab{a}}, eprint arXiv,
  1111, 5019

\bibitem[{Bonfils {et~al.}(2011{\natexlab{b}})Bonfils, Gillon, Forveille,
  Delfosse, Deming, Demory, Lovis, Mayor, Neves, Perrier, Santos, Seager, Udry,
  Boisse, \& Bonnefoy}]{Bonfils:2011a}
Bonfils, X., Gillon, M., Forveille, T., {et~al.} 2011{\natexlab{b}}, \aap, 528,
  111

\bibitem[{Borucki {et~al.}(2011)Borucki, Koch, Basri, Batalha, Brown, Bryson,
  Caldwell, Christensen-Dalsgaard, Cochran, DeVore, Dunham, Gautier, Geary,
  Gilliland, Gould, Howell, Jenkins, Latham, Lissauer, Marcy, Rowe, Sasselov,
  Boss, Charbonneau, Ciardi, Doyle, Dupree, Ford, Fortney, Holman, Seager,
  Steffen, Tarter, Welsh, Allen, Buchhave, Christiansen, Clarke, D{\'e}sert,
  Endl, Fabrycky, Fressin, Haas, Horch, Howard, Isaacson, Kjeldsen,
  Kolodziejczak, Kulesa, Li, Machalek, McCarthy, MacQueen, Meibom, Miquel,
  Prsa, Quinn, Quintana, Ragozzine, Sherry, Shporer, Tenenbaum, Torres,
  Twicken, Cleve, \& Walkowicz}]{Borucki:2011}
Borucki, W.~J., Koch, D.~G., Basri, G., {et~al.} 2011, eprint arXiv, 1102, 541

\bibitem[{Butler {et~al.}(2004)Butler, Vogt, Marcy, Fischer, Wright, Henry,
  Laughlin, \& Lissauer}]{Butler:2004}
Butler, R.~P., Vogt, S.~S., Marcy, G.~W., {et~al.} 2004, \apj, 617, 580

\bibitem[{Charbonneau {et~al.}(2009)Charbonneau, Berta, Irwin, Burke, Nutzman,
  Buchhave, Lovis, Bonfils, Latham, Udry, Murray-Clay, Holman, Falco, Winn,
  Queloz, Pepe, Mayor, Delfosse, \& Forveille}]{Charbonneau:2009}
Charbonneau, D., Berta, Z.~K., Irwin, J., {et~al.} 2009, \nat, 462, 891

\bibitem[{Claret(2000)}]{Claret:2000}
Claret, A. 2000, \aap, 363, 1081

\bibitem[{Cutri {et~al.}(2003)Cutri, Skrutskie, van Dyk, Beichman, Carpenter,
  Chester, Cambresy, Evans, Fowler, Gizis, Howard, Huchra, Jarrett, Kopan,
  Kirkpatrick, Light, Marsh, McCallon, Schneider, Stiening, Sykes, Weinberg,
  Wheaton, Wheelock, \& Zacarias}]{Cutri:2003}
Cutri, R.~M., Skrutskie, M.~F., van Dyk, S., {et~al.} 2003, The IRSA 2MASS
  All-Sky Point Source Catalog

\bibitem[{Delfosse {et~al.}(2000)Delfosse, Forveille, S{\'e}gransan, Beuzit,
  Udry, Perrier, \& Mayor}]{Delfosse:2000}
Delfosse, X., Forveille, T., S{\'e}gransan, D., {et~al.} 2000, \aap, 364, 217

\bibitem[{Delorme {et~al.}(2011)Delorme, Cameron, Hebb, Rostron, Lister,
  Norton, Pollacco, \& West}]{Delorme:2011}
Delorme, P., Cameron, A.~C., Hebb, L., {et~al.} 2011, \mnras, 413, 2218

\bibitem[{Demory {et~al.}(2011)Demory, Gillon, Deming, Valencia, Seager,
  Benneke, Lovis, Cubillos, Harrington, Stevenson, Mayor, Pepe, Queloz,
  Segransan, \& Udry}]{Demory:2011}
Demory, B.~O., Gillon, M., Deming, D., {et~al.} 2011, eprint arXiv, 1105, 415,
  submitted to A{\&}A

\bibitem[{Demory {et~al.}(2009)Demory, S{\'e}gransan, Forveille, Queloz,
  Beuzit, Delfosse, di~Folco, Kervella, Bouquin, Perrier, Benisty, Duvert,
  Hofmann, Lopez, \& Petrov}]{Demory:2009}
Demory, B.-O., S{\'e}gransan, D., Forveille, T., {et~al.} 2009, \aap, 505, 205

\bibitem[{Eggen(1990)}]{Eggen:1990}
Eggen, O.~J. 1990, Astronomical Society of the Pacific, 102, 166

\bibitem[{Figueira {et~al.}(2012)Figueira, Marmier, Bou{\'e}, Lovis, Santos,
  Montalto, Udry, Pepe, \& Mayor}]{Figueira:2012}
Figueira, P., Marmier, M., Bou{\'e}, G., {et~al.} 2012, eprint arXiv, 1202,
  2801

\bibitem[{Ford(2005)}]{Ford:2005}
Ford, E.~B. 2005, \aj, 129, 1706

\bibitem[{Fortney {et~al.}(2007)Fortney, Marley, \& Barnes}]{Fortney:2007}
Fortney, J.~J., Marley, M.~S., \& Barnes, J.~W. 2007, \apj, 659, 1661

\bibitem[{Fressin {et~al.}(2012)Fressin, Torres, Rowe, Charbonneau, Rogers,
  Ballard, Batalha, Borucki, Bryson, Buchhave, Ciardi, D{\'e}sert, Dressing,
  Fabrycky, Ford, Gautier, Henze, Holman, Howard, Howell, Jenkins, Koch,
  Latham, Lissauer, Marcy, Quinn, Ragozzine, Sasselov, Seager, Barclay,
  Mullally, Seader, Still, Twicken, Thompson, \& Uddin}]{Fressin:2012}
Fressin, F., Torres, G., Rowe, J.~F., {et~al.} 2012, \nat, 482, 195, (c) 2012:
  Nature

\bibitem[{Gautier {et~al.}(2012)Gautier, Charbonneau, Rowe, Marcy, Isaacson,
  Torres, Fressin, Rogers, D{\'e}sert, Buchhave, Latham, Quinn, Ciardi,
  Fabrycky, Ford, Gilliland, Walkowicz, Bryson, Cochran, Endl, Fischer, Howell,
  Horch, Barclay, Batalha, Borucki, Christiansen, Geary, Henze, Holman,
  Ibrahim, Jenkins, Kinemuchi, Koch, Lissauer, Sanderfer, Sasselov, Seager,
  Silverio, Smith, Still, Stumpe, Tenenbaum, \& Cleve}]{Gautier:2012}
Gautier, T.~N., Charbonneau, D., Rowe, J.~F., {et~al.} 2012, \apj, 749, 15

\bibitem[{Gillon {et~al.}(2007{\natexlab{a}})Gillon, Demory, Barman, Bonfils,
  Mazeh, Pont, Udry, Mayor, \& Queloz}]{Gillon:2007b}
Gillon, M., Demory, B.-O., Barman, T., {et~al.} 2007{\natexlab{a}}, \aap, 471,
  L51

\bibitem[{Gillon {et~al.}(2012)Gillon, Demory, Benneke, Valencia, Deming,
  Seager, Lovis, Mayor, Pepe, Queloz, S{\'e}gransan, \& Udry}]{Gillon:2012}
Gillon, M., Demory, B.-O., Benneke, B., {et~al.} 2012, \aap, 539, 28

\bibitem[{Gillon {et~al.}(2011)Gillon, Doyle, Lendl, Maxted, Triaud, Anderson,
  Barros, Bento, Collier-Cameron, Enoch, Faedi, Hellier, Jehin, Magain,
  Montalb{\'a}n, Pepe, Pollacco, Queloz, Smalley, Segransan, Smith, Southworth,
  Udry, West, \& Wheatley}]{Gillon:2011}
Gillon, M., Doyle, A.~P., Lendl, M., {et~al.} 2011, \aap, 533, 88

\bibitem[{Gillon {et~al.}(2007{\natexlab{b}})Gillon, Pont, Demory, Mallmann,
  Mayor, Mazeh, Queloz, Shporer, Udry, \& Vuissoz}]{Gillon:2007a}
Gillon, M., Pont, F., Demory, B.-O., {et~al.} 2007{\natexlab{b}}, \aap, 472,
  L13

\bibitem[{Gregory(2005)}]{Gregory:2005}
Gregory, P.~C. 2005, \apj, 631, 1198

\bibitem[{Gregory(2007)}]{Gregory:2007}
Gregory, P.~C. 2007, \mnras, 374, 1321

\bibitem[{Hartman {et~al.}(2011)Hartman, Bakos, Kipping, Torres, Kov{\'a}cs,
  Noyes, Latham, Howard, Fischer, Johnson, Marcy, Isaacson, Quinn, Buchhave,
  B{\'e}ky, Sasselov, Stefanik, Esquerdo, Everett, Perumpilly, L{\'a}z{\'a}r,
  Papp, \& S{\'a}ri}]{Hartman:2011c}
Hartman, J.~D., Bakos, G.~{\'A}., Kipping, D.~M., {et~al.} 2011, \apj, 728, 138

\bibitem[{Haywood {et~al.}(1997)Haywood, Robin, \& Creze}]{Haywood:1997}
Haywood, M., Robin, A.~C., \& Creze, M. 1997, \aap, 320, 428

\bibitem[{Henry {et~al.}(2011)Henry, Howard, Marcy, Fischer, \&
  Johnson}]{Henry:2011}
Henry, G.~W., Howard, A.~W., Marcy, G.~W., Fischer, D.~A., \& Johnson, J.~A.
  2011, eprint arXiv, 1109, 2549, 8 pages, 7 figures, submitted to ApJ

\bibitem[{Jehin {et~al.}(2011)Jehin, Gillon, Queloz, Magain, Manfroid, Chantry,
  Lendl, Hutsem{\'e}kers, \& Udry}]{Jehin:2011}
Jehin, E., Gillon, M., Queloz, D., {et~al.} 2011, The Messenger, 145, 2

\bibitem[{Leggett(1992)}]{Leggett:1992}
Leggett, S.~K. 1992, \apjs, 82, 351

\bibitem[{Leggett {et~al.}(2001)Leggett, Allard, Geballe, Hauschildt, \&
  Schweitzer}]{Leggett:2001}
Leggett, S.~K., Allard, F., Geballe, T.~R., Hauschildt, P.~H., \& Schweitzer,
  A. 2001, \apj, 548, 908

\bibitem[{Lendl {et~al.}(2012)Lendl, Anderson, Collier-Cameron, Doyle, Gillon,
  Hellier, Jehin, Lister, Maxted, Pepe, Pollacco, Queloz, Smalley, Segransan,
  Smith, Triaud, Udry, West, \& Wheatley}]{Lendl:2012}
Lendl, M., Anderson, D.~R., Collier-Cameron, A., {et~al.} 2012, eprint arXiv,
  1205, 2757

\bibitem[{L{\'e}pine(2005)}]{Lepine:2005a}
L{\'e}pine, S. 2005, \aj, 130, 1680

\bibitem[{L{\'e}pine \& Shara(2005)}]{Lepine:2005b}
L{\'e}pine, S. \& Shara, M.~M. 2005, \aj, 129, 1483

\bibitem[{Lissauer {et~al.}(2011{\natexlab{a}})Lissauer, Fabrycky, Ford,
  Borucki, Fressin, Marcy, Orosz, Rowe, Torres, Welsh, Batalha, Bryson,
  Buchhave, Caldwell, Carter, Charbonneau, Christiansen, Cochran, Desert,
  Dunham, Fanelli, Fortney, Gautier, Geary, Gilliland, Haas, Hall, Holman,
  Koch, Latham, Lopez, McCauliff, Miller, Morehead, Quintana, Ragozzine,
  Sasselov, Short, \& Steffen}]{Lissauer:2011b}
Lissauer, J.~J., Fabrycky, D.~C., Ford, E.~B., {et~al.} 2011{\natexlab{a}},
  \nat, 470, 53, (c) 2011: Nature

\bibitem[{Lissauer {et~al.}(2011{\natexlab{b}})Lissauer, Ragozzine, Fabrycky,
  Steffen, Ford, Jenkins, Shporer, Holman, Rowe, Quintana, Batalha, Borucki,
  Bryson, Caldwell, Carter, Ciardi, Dunham, Fortney, Gautier, Howell, Koch,
  Latham, Marcy, Morehead, \& Sasselov}]{Lissauer:2011a}
Lissauer, J.~J., Ragozzine, D., Fabrycky, D.~C., {et~al.} 2011{\natexlab{b}},
  The Astrophysical Journal Supplement, 197, 8

\bibitem[{Mayor {et~al.}(2003)Mayor, Pepe, Queloz, Bouchy, Rupprecht, Curto,
  Avila, Benz, Bertaux, Bonfils, Dall, Dekker, Delabre, Eckert, Fleury,
  Gilliotte, Gojak, Guzman, Kohler, Lizon, Longinotti, Lovis, Megevand,
  Pasquini, Reyes, Sivan, Sosnowska, Soto, Udry, van Kesteren, Weber, \&
  Weilenmann}]{Mayor:2003}
Mayor, M., Pepe, F., Queloz, D., {et~al.} 2003, The Messenger, 114, 20

\bibitem[{McCormac(2012)}]{McCormac:2012}
McCormac. 2012

\bibitem[{Muirhead {et~al.}(2012)Muirhead, Johnson, Apps, Carter, Morton,
  Fabrycky, Pineda, Bottom, Rojas-Ayala, Schlawin, Hamren, Covey, Crepp,
  Stassun, Pepper, Hebb, Kirby, Howard, Isaacson, Marcy, Levitan, Diaz-Santos,
  Armus, \& Lloyd}]{Muirhead:2012}
Muirhead, P.~S., Johnson, J.~A., Apps, K., {et~al.} 2012, \apj, 747, 144

\bibitem[{Nutzman \& Charbonneau(2008)}]{Nutzman:2008}
Nutzman, P. \& Charbonneau, D. 2008, \pasp, 120, 317

\bibitem[{Pepe {et~al.}(2004)Pepe, Mayor, Queloz, Benz, Bonfils, Bouchy, Curto,
  Lovis, M{\'e}gevand, Moutou, Naef, Rupprecht, Santos, Sivan, Sosnowska, \&
  Udry}]{Pepe:2004}
Pepe, F., Mayor, M., Queloz, D., {et~al.} 2004, \aap, 423, 385

\bibitem[{Press {et~al.}(1992)Press, Teukolsky, Vetterling, \&
  Flannery}]{Press:1992}
Press, W.~H., Teukolsky, S.~A., Vetterling, W.~T., \& Flannery, B.~P. 1992,
  Cambridge: University Press

\bibitem[{Rapaport {et~al.}(2001)Rapaport, Campion, Soubiran, Daigne,
  P{\'e}ri{\'e}, Bosq, Colin, Desbats, Ducourant, Mazurier, Montignac, Ralite,
  R{\'e}qui{\`e}me, \& Viateau}]{Rapaport:2001}
Rapaport, M., Campion, J.-F.~L., Soubiran, C., {et~al.} 2001, \aap, 376, 325

\bibitem[{Reid {et~al.}(1997)Reid, Hawley, \& Gizis}]{Reid:1997}
Reid, I.~N., Hawley, S.~L., \& Gizis, J.~E. 1997, VizieR On-line Data Catalog,
  3198, 0

\bibitem[{Seager \& Mall{\'e}n-Ornelas(2003)}]{Seager:2003}
Seager, S. \& Mall{\'e}n-Ornelas, G. 2003, \apj, 585, 1038

\bibitem[{Winn {et~al.}(2011)Winn, Matthews, Dawson, Fabrycky, Holman,
  Kallinger, Kuschnig, Sasselov, Dragomir, Guenther, Moffat, Rowe, Rucinski, \&
  Weiss}]{Winn:2011}
Winn, J.~N., Matthews, J.~M., Dawson, R.~I., {et~al.} 2011, eprint arXiv, 1104,
  5230, submitted to ApJ Letters

\bibitem[{Zechmeister \& K{\"u}rster(2009)}]{Zechmeister:2009b}
Zechmeister, M. \& K{\"u}rster, M. 2009, \aap, 496, 577

\end{thebibliography}

\end{document}